\documentclass[prb,twocolumn,aps,amsmath,showpacs]{revtex4}
\usepackage{graphicx,bm,times,color}
 
\begin{document}

\title{
Correlation function for the one-dimensional extended Hubbard model
 at quarter filling
}

\author{S. Nishimoto}
\affiliation{Leibniz-Institut f\"ur Festk\"orper- und Werkstoffforschung 
Dresden, D-01171 Dresden, Germany}

\author{M. Tsuchiizu}
\affiliation{Department of Physics, Nagoya University,
         Nagoya 464-8602, Japan}
 
\date{\today}
 
\begin{abstract}
We examine the density-density correlation function in the 
  Tomonaga-Luttinger liquid 
  state for the one-dimensional extended Hubbard model
 with the on-site Coulomb repulsion $U$ and the
  intersite repulsion $V$ at quarter filling.
By taking into account the effect of the 
  marginally irrelevant umklapp scattering operator
  by utilizing the renormalization-group technique based on 
  the bosonization method,
  we obtain the generalized analytical form of the correlation function.
We show that, in the proximity to the gapped charge-ordered phase,  
  the correlation function 
  exhibits anomalous crossover  between the pure power-law behavior and 
the power-law behavior with logarithmic corrections, 
   depending on the length scale.
Such a crossover is also confirmed by 
  the highly-accurate numerical density-matrix renormalization group method.
\end{abstract}

\pacs{71.10.Fd, 71.10.Hf, 71.10.Pm}

\maketitle

\section{Introduction}

One-dimensional (1D) electron and spin systems have been attracted  
  much attention since they often exhibit nontrivial 
  quasi-long-range-ordered behavior 
 due to the large low-dimensional quantum fluctuation effects.
  \cite{Gogolin_book,Giamarchi_book}
The critical behavior in the 1D systems, which is called the 
 Tomonaga-Luttinger liquid (TLL) state, 
 has a long history of research and
  the low-lying modes are known to be described by 
  collective gapless excitations and 
  physical quantities show power-law behavior 
 in the temperature and/or distance dependences.
It has also been recognized that, 
 in the systems with spin-rotational symmetry, 
 logarithmic singularities appear in
 the magnetic-field-dependent 
corrections to the magnetization \cite{YangYang1966,Babujian1983} 
  and the spin 
susceptibility, \cite{Schlottmann1987} 
  and in the temperature-dependent corrections 
to the spin susceptibility, \cite{Eggert1994} specific heat, \cite{Kluemper1998} 
and nuclear magnetic resonance, \cite{Barzykin2000} etc. 
Motivated by such 
developments in the field, a number of numerical studies has been successfully 
performed to examine the logarithmic corrections
  in spin-chain systems.
\cite{Kubo1988,Liang1990,Sandvik1993,Hallberg1995,Koma1996,Eggert1996,Nomura1993,Hikihara1998}
The spin-spin correlation function of the 1D $S=\frac{1}{2}$ Heisenberg model 
has been extensively studied 
  as a most fundamental theoretical model to examine
 the presence of logarithmic corrections. 
 \cite{Finkel'stein1997,Affleck1989,Giamarchi1989,Singh1989} 
At this moment, the correlation amplitudes in the asymptotic form of 
the correlation function can be exactly obtained. \cite{Lukyanov1997,Lukyanov1999,Lukyanov2003} 
In contrast, only few efforts have been devoted to those of 
the Hubbard model due in part to the difficulty of analyzing. 
\cite{Schulz1990}
Therefore, 
the situation is much less satisfactory as far as the logarithmic corrections 
in the Hubbard model are concerned.

In the present paper, we focus on the logarithmic corrections 
  in the equal-time density-density correlation function 
  of the quarter-filled Hubbard model
  including the Coulomb repulsion between electrons on 
  site $U$ and 
  the nearest-neighbor sites $V$. 
So far this model 
 has been analyzed 
  as  a minimal model  to describe  physical phenomena
  in  organic solids. \cite{Seo_review} 
The Hamiltonian of 
the 1D extended Hubbard model at quarter-filling 
 is given by 
\begin{eqnarray}
H &=&
-t\sum_{j,s}
\left(
c_{j,s}^\dagger c_{j+1,s}^{} + \mathrm{H.c.}
\right)
\nonumber \\
&& {} +U \sum_j n_{j,\uparrow} n_{j,\downarrow}
+V\sum_j n_{j} n_{j+1},
\label{eq:EHM}
\end{eqnarray}
where $c_{j,s}$ is the annihilation operator on the $j$th site with 
  spin $s(=\uparrow,\downarrow)$, and the density operators are
$n_j=n_{j,\uparrow}+n_{j,\downarrow}$ and
$n_{j,s}\equiv \, :  c_{j,s}^\dagger \, c_{j,s}^{} : \, 
  = c_{j,s}^\dagger \, c_{j,s}^{} - \frac{1}{4}$.
The hopping energy between the nearest-neighbor sites 
  is represented by $t$.
It is known that, at zero temperature,
  the gapped charge-ordered (CO) state emerges 
  in the large repulsive $U$ and $V$ region, 
where 
  the phase boundary 
  is  determined numerically
  by using the  exact diagonalization 
  \cite{Mila1993,Sano1994} and 
  the highly-accurate density-matrix
 renormalization group (DMRG) method.  \cite{Ejima2005}
The mechanism of this quantum phase transition has also been
  addressed by the bosonization 
  technique and the renormalization-group (RG) method.
\cite{Giamarchi_book,Schulz1994,Tsuchiizu2001}
In this paper, we perform the detailed analysis on  
  the equal-time correlation function 
  $N(x)\equiv\langle n_j n_{j+x} \rangle$
  in the TLL phase.
The exponent of the 
 correlation functions 
  is characterized by so-called the
   TLL parameter $K_\rho$.  \cite{Schulz1990} 
Especially by focusing on the correlation function near the boundary 
  to the CO insulating state,
  we show that it exhibits
 the nontrivial crossover, depending on the length scale,  from the 
  power-law behavior with logarithmic correction for short distance,
  to the pure power-law behavior for large distance. 

The present paper is organized as follows.
In Sec.\ II, the analytical form of the 
 correlation function is obtained by utilizing 
 the RG technique based on the 
  bosonization method.
In Sec.\ III, the analytical results are confirmed by using 
  the highly-accurate DMRG
  method.
The summary is given in Sec.\ IV.
Detailed derivation of the analytical form of the
  correlation function is given in the Appendix.

\section{Bosonization approach}

In this section, 
we derive the generalized analytical form of the correlation function
  in the TLL state.
We analyze the $U\to \infty $ limit case and the finite $U$ case
  separately, since the picture of the 
  $U\to \infty$ can become transparent with the analogy of the
  spin-chain system which properties are well understood.

\subsection{The $U\to \infty$ limit}

In the $U\to \infty$ limit, since
  the double occupancy of electrons is excluded,
the extended Hubbard model [Eq.\ (\ref{eq:EHM})] reduces to the 
  \textit{spinless half-filled} model:
\begin{eqnarray}
H_{U\to \infty} =
-t\sum_{j}
\left(
d_{j}^\dagger d_{j+1}^{} + \mathrm{H.c.}
\right)
+V\sum_j n^d_{j} n^d_{j+1},
\label{eq:spinlessmodel}
\end{eqnarray}
where $n^d_j=d_j^\dagger d_j^{}-\frac{1}{2}$.
It is well known that 
 this model can be mapped onto the XXZ spin-chain 
  model by using the Jordan-Wigner transformation
 \cite{Giamarchi_book,Affleck}
  and the physical properties have been extensively studied 
  with both the exact treatment based on the Bethe ansatz 
  and numerical approaches.
Here we examine the analytical form of the correlation function 
  by using the exact results obtained in the context of 
  spin-chain problems.

The  density operator is
 expressed in terms of the bosonic field operator $\varphi$ as 
\cite{Giamarchi_book,Affleck,Hikihara1998}
\begin{eqnarray}
\rho(x) 
=
\frac{n^d_j}{a}
=
\frac{1}{2\pi \eta}  \frac{d \varphi}{dx}
  - (-1)^j \frac{c}{\pi a} \sin \sqrt{\frac{2\pi}{\eta}} \, \varphi,
\label{eq:bosonization-density}
\end{eqnarray}
where $a$ is the lattice constant and we will set $a=1$ in the
  following.
The parameter $\eta$ can be related to the TLL parameter 
by $K_\rho\equiv 1/(4\eta)$.
The nonuniversal parameter $c$ will be shown later.
The model Hamiltonian (\ref{eq:spinlessmodel}) can be expressed
  in terms of the bosonic field $\varphi$.

It is well-known that 
the TLL phase is realized for small $V \le V_c(= 2t)$, while
the gapped CO state appears for $V>V_c$.
In the TLL phase, 
the parameter $\eta$ is known exactly from the Bethe ansatz as
  $\eta \equiv  1 - \pi^{-1} \cos^{-1} (V/2t)$,
  and the TLL parameter $K_\rho = 1/(4\eta)$ varies within the range
  of $1/4\le K_\rho \le 1/2$ for $0 \le V \le V_c$.
The TLL parameter $K_\rho$ approaches to 
  an universal value $K_\rho=1/4$ when $V\to V_c$.
The mechanism of this quantum phase transition has also been
  addressed by the bosonization 
  technique and the RG method 
\cite{Giamarchi_book,Schulz1994,Tsuchiizu2001}
  and it has been clarified that 
  the 1/4-filled umklapp scattering has a crucial role 
  in making the TLL state into the gapped CO state.
It has also been shown that 
  the universality class of this TLL-to-CO phase transition 
  is in the Kosterlitz-Thouless transition, where
  the umklapp scattering is irrelevant in the TLL phase while 
  it becomes relevant in the CO phase.
It is worthwhile to note that, on the phase boundary between the TLL
  and CO states, 
  the umklapp scattering term becomes \textit{marginally irrelevant},
  where it shows very slow scaling-parameter dependence
  and can give rise to anomalous corrections to physical quantities.
We analyze this effect on 
 the density-density correlation function 
  by using the RG method developed in Ref.\ 
  \onlinecite{Giamarchi1989}.
The resultant form of the correlation function 
 for $U=\infty$ in the TLL state is given by
\begin{equation}
N(x)
=
- \frac{K_\rho}{\pi^2 x^2} 
+ A_2  \frac{\cos 4k_Fx}{x^{4K_\rho}} 
\frac{\left[ 1- (\alpha/x)^{2-8K_\rho}\right]^{1/2}}
     {\left[ 1+ (\alpha/x)^{2-8K_\rho}\right]^{3/2}} ,
\label{eq:generalformUinf}
\end{equation}
where $k_F$ is the Fermi wave vector, $k_F(=\pi/4)$.
The derivation of this form is given in the Appendix.
The parameter $\alpha$ is the short-distance cutoff which appears 
  in the RG method.
The coefficient $A_2$ is given by  $A_2\equiv c^2/(2\pi^2)$
  by using the parameter $c$ in Eq.\ (\ref{eq:bosonization-density}).
In Refs.\ \onlinecite{Lukyanov1997} and \onlinecite{Lukyanov1999}, 
 the exact form of the parameter $c$ has been proposed, where
it is controlled by the TLL parameter, i.e., $c=c(\eta)$, and 
  its exact form is given by 
\begin{widetext}
\begin{eqnarray}
c(\eta)&=&
2
\left[
  \frac{\Gamma\biglb( \eta/(2-2\eta)\bigrb)}
       {2\sqrt{\pi} \, \Gamma \biglb( 1/(2-2\eta)\bigrb)}
\right]^{1/2\eta}
\exp
\left[
 \frac{1}{2} \int_0^\infty \frac{dt}{t}
   \left(
      \frac{\sinh[(2\eta-1)t]}{\sinh(\eta t)\, \cosh[(1-\eta)t]}
  - \frac{2\eta-1}{\eta} e^{-2t}
   \right)
\right].
\label{eq:c}
\end{eqnarray}
\end{widetext}
The numerical values of $A_2$ are shown in Table.\ I. 
For $(U,V)=(\infty,0)$, the quantity  $c(\eta)$ becomes $c(\eta=1/2)=1$.
We note here that 
  the quantity $\alpha$ in Eq.\ (\ref{eq:generalformUinf})
  is the only unknown parameter,
  which is to be determined numerically.

\begin{table*}[t]
\caption{
Amplitude $A_2=c^2(\eta)/(2\pi^2)$ 
 as a functions of $V/t$ for $U=\infty$.
}
\begin{ruledtabular}
\begin{tabular}{cc|cc|cc|cc|cc}
  $V/t$  &   $A_2$    &  $V/t$  &   $A_2$  &  
  $V/t$  &   $A_2$    &  $V/t$  &   $A_2$  &  $V/t$  &   $A_2$  \\
\hline
0.0      & 0.0506606  &0.5      & 0.0741643  &
  1.0  &  0.107134  &  1.5  &  0.164769  &
1.95   &   0.404613\\
0.1      & 0.0548623  &0.6      & 0.0797783  &
  1.1  &  0.115734  &  1.6  &  0.184141  &
1.99   &   0.655754\\
0.2      & 0.0592869  &0.7      & 0.0858022  &
  1.2  &  0.125393  &  1.7  &  0.209986  &
1.995  &   0.796158\\
0.3      & 0.0639583  &0.8      & 0.0923055  &
  1.3  &  0.136425  &  1.8  &  0.248441  &
1.999  &   1.22703\\
0.4      & 0.0689056  &0.9      & 0.0993773  &
  1.4  &  0.149299  &  1.9  &  0.321076  &
1.99999  &  3.98564     
\end{tabular}
\end{ruledtabular}
\end{table*}

Here we find that equation (\ref{eq:generalformUinf})
 has two different asymptotics:
(i) In the short-range region, 
the power-law behavior \textit{with logarithmic correction} is obtained, 
   while (ii)
  the logarithmic correction disappears in
  the long-range region.
By noting $(x/\alpha)^{-(8K_\rho-2)}=\exp[-(8K_\rho-2)\ln(x/\alpha)]$,
the length scale $x_{\mathrm{cross}}$
  which characterizes crossover between these two
  regions is given by
\begin{eqnarray}
x_{\mathrm{cross}} = \alpha \exp[1/(8K_\rho-2)].
\end{eqnarray}
We note that, for $V= V_c(= 2t)$, 
the logarithmic correction appears in the whole length scale.
By noting 
$A_2\to  (2-V/t)^{-1/4}/(\sqrt{2}\pi)$ for $V\to V_c$, 
 the explicit form of 
 the correlation function (\ref{eq:generalformUinf}) 
  at $V=2t$ is given by
\begin{eqnarray}
N(x)&=& - \frac{1}{4\pi^2 x^2} 
  +  \frac{1}{(2\pi)^{3/2}} \frac{\cos 4k_Fx}{x} \ln^{1/2} (x/\alpha)
\nonumber \\
&& \hspace*{2cm}
\mbox{for } (U,V)=(\infty,2t).
\end{eqnarray}
This formula was
  reported in Ref.\ \onlinecite{Affleck1998} 
  for the $S=\frac{1}{2}$ antiferromagnetic Heisenberg spin chain.

\subsection{For finite $U$}

Next we examine the generic $0<U\neq \infty$ case.
In this case,
there appears the conventional $2k_F$ oscillation term
  in addition to the $4k_F$ one.
We show that
  the additional logarithmic correction 
   appears near the phase boundary to the CO phase,
    not only in the $4k_F$ oscillation  term but in 
   the $2k_F$ oscillation term.

Based on the conventional bosonization 
  for electron systems, 
the density operator is given by
\cite{Giamarchi_book,Schulz1994}
\begin{eqnarray}
\rho(x) &=& \frac{1}{\pi}  \frac{d \phi_{\rho}}{dx}
  - \frac{2c_1}{\pi a} \sin (2k_F x + \phi_\rho ) \, \cos \phi_\sigma
\nonumber \\
&&
  + \frac{2c_2}{\pi a} \cos (4k_F x + 2\phi_\rho ) ,
\end{eqnarray}
where  $\phi_\rho$ and $\phi_\sigma$ are the charge and spin phase 
  fields.
The $c_1$ and $c_2$ are nonuniversal numerical quantities 
 satisfying $c_1=1$ and $c_2=0$ in the noninteracting case.
In the similar way to the $U=\infty$ case,
the most general form of the density-density correlation function 
 is derived as (see Appendix)
\begin{eqnarray}
N(x)
&=& - \frac{K_\rho}{\pi^2 x^2} 
+ A_1 \frac{\cos 2k_Fx}{x^{K_\rho+1}}
\frac{\ln^{-3/2} (x/\alpha_\sigma)}
     {\left[ 1- (\alpha/x)^{4-16K_\rho}\right]^{1/8}}
\nonumber \\ && {}
+ A_2  \frac{\cos 4k_Fx}{x^{4K_\rho}} 
\frac{\left[ 1- (\alpha/x)^{2-8K_\rho}\right]^{1/2}}
     {\left[ 1+ (\alpha/x)^{2-8K_\rho}\right]^{3/2}} ,
\label{eq:generalform}
\end{eqnarray}
where $\alpha$ and $\alpha_\sigma$ are  the short-distance
 cutoff parameters for the charge and spin sectors, respectively.
In the case of $U\neq \infty$,
 the coefficients $A_1$ and $A_2$, which are proportional to 
  $c_1^2$ and $c_2^2$ respectively,
  are to be determined numerically. 
The logarithmic correction 
  $\ln^{-3/2} (x/\alpha_\sigma)$ in the $2k_F$
  oscillating term appears due to the 
  marginally irrelevant coupling of the spin channel.
  \cite{Giamarchi1989}
In the noninteracting limit $U=V=0$, the quantity $A_2$ 
  vanishes and the logarithmic correction 
  $\ln^{-3/2} (x/\alpha_\sigma)$ is replaced by a constant, and
 then the correlation function reproduces the 
  trivial result 
  $N(x)= - 1/(\pi^2 x^2) + \cos 2k_Fx/(\pi^2 x^2)$.

From Eq.\ (\ref{eq:generalform}), we find that 
  an anomalous logarithmic correction also appears in the $2k_F$ oscillating 
  term near the boundary of the CO phase.
On the phase boundary, the correlation function reads
\begin{eqnarray}
N(x)&=& - \frac{1}{4\pi^2 x^2}
\nonumber \\  && {}
 + \tilde A_1 \frac{\cos 2k_F x}{x^{5/4}} \ln^{-3/2}(x/\alpha_\sigma)
 \ln^{-1/8}(x/\alpha)
\nonumber \\ && {}
 + \tilde A_2 \frac{\cos 4k_Fx}{(2\pi)^{3/2}} \ln^{1/2} (x/\alpha)
\nonumber \\
&& \hspace*{2cm}
\mbox{for } (U,V)=(U_c,V_c)
\end{eqnarray}
where 
$\tilde A_1 \to 0$ and $\tilde A_2 \to 1$ 
for $(U_c,V_c)\to(\infty,2)$.

\section{Numerical results}

For numerical confirmation of the logarithmic corrections, 
we employ the DMRG technique which provides very accurate data for 
the ground-state correlation functions of 1D 
correlated electron systems.~\cite{White1992} 
We consider $L/2$ electrons on a chain with $L$ sites and 
calculate the equal-time density-density correlation function
\begin{equation}
N(x)=\langle n_j n_{j+x} \rangle -\langle n_j \rangle \langle n_{j+x} \rangle,
\label{Nx}
\end{equation}
under the open-end boundary conditions (OBC). Here, the distance $x$ 
is centered at the middle of the system. The application of OBC enables 
us to obtain the correlation function (\ref{Nx}) quite accurately 
for very large finite-size systems up to $\sim {\cal O}(1000)$ sites. 
However, the real-space DMRG method works with finite number of sites, 
so that we have to pay special attention to the finite-size effects for 
a precise comparison with the RG results. In the present calculations, 
the most problematic finite-size effect is the Friedel oscillation starting 
from the open edges. To eliminate it, we simply add on-site potential energy 
$V/2$ on both edge sites. It corresponds to a compensation of the ``missing 
correlation'' caused by the absence of their neighboring site. Hereby 
the Friedel oscillation is fairly suppressed. 

On that basis, the remaining finite-size effects are investigated. We now 
choose some parameters in the vicinity of the CO phase where the finite-size 
effect is relatively large due to strong charge fluctuations. 
For these parameters, we calculate $N(x)$ for several chains with length 
up to $L=1024$ sites and then obtain the extrapolated values to 
the thermodynamic limit ($L \to \infty$) using the finite-size-scaling analysis. 
By comparing the extrapolated values and the finite-size data, we find 
that $N(x \le 200)$ in the thermodynamic limit can be reproduced with 
extracted central $200$ sites of a chain with $L=512$ sites within 
a few percent error. The relative error in the ground-state energy, 
$|\{e(\infty)-e(512)\}/e(\infty)|$, is below $0.1\%$, where $e(L)$ is 
the ground-state energy per site for a chain with $L$ sites. 
Consequently, we will study the equal-time correlation function $N(x)$ 
for the central $200$ sites of a chain with $L=512$ sites without 
the finite-size-scaling analysis. We keep up to $m \approx 4000$ density-matrix 
eigenstates in the DMRG procedure and all the calculated quantities 
are extrapolated to the $m \to \infty$ limit. For your information, 
in this way we obtain $(A_1, A_2)=(0.991,-0.0003)$ [the exact are 
$(A_1, A_2)=(1,0)$] for the coefficients of Eq.\ (\ref{eq:generalform}) 
in the non-interacting case $U=V=0$ which poses a non-trivial problem 
to the DMRG method.

\subsection{The $U\to \infty$ limit}

\begin{figure}[t]
\includegraphics[width=6.0cm]{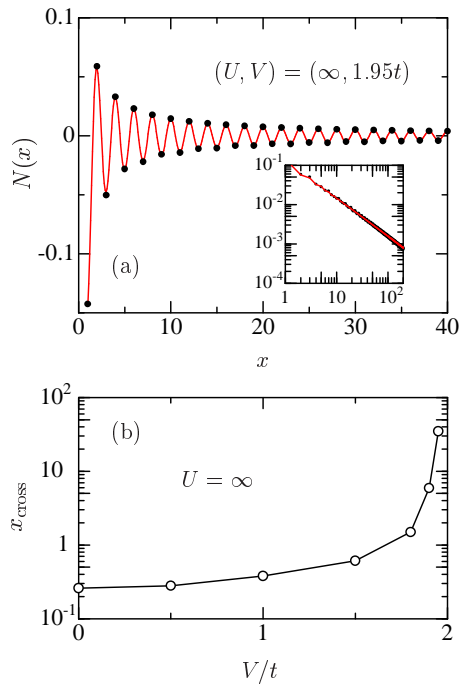}
\caption{
(Color online)
(a) DMRG results of the correlation function $N(x)$ for $(U,V)=(\infty, 1.95t)$. 
The solid line denotes a fitting with Eq.\ (\ref{eq:generalformUinf}). 
Inset: $|N(x)|$ plotted on a log-log scale. 
(b) Estimated length scale of the logarithmic correction $x_{\mathrm{cross}}$ as 
a function of $V/t$ for $U=\infty$.
\label{fig:fit_17_V195}
}
\end{figure}

Let us first consider the correlation function $N(x)$ in the $U \to \infty$ limit. 
We now attempt to fit the DMRG results of $N(x)$ into the analytical form of 
Eq.\ (\ref{eq:generalformUinf}). Since the exact solutions of $A_2$ and $K_\rho$ 
are available, the quantity $\alpha$ is the only fitting parameter in 
Eq.\ (\ref{eq:generalformUinf}). Figure~\ref{fig:fit_17_V195}\ (a) shows the DMRG 
results of $N(x)$ for $(U,V)=(\infty, 1.95t)$. An excellent agreement of the DMRG 
data with the fitted line is found. We then obtain $\alpha=0.0515$, which leads to 
$x_{\mathrm{cross}}=34.8$. It means that the logarithmic correction appears at 
$x \lesssim 35$ for $V=1.95t$. We note that the central $200$ sites out of $L=512$ 
are used to carry out the fitting procedure and, however, the fitting results are 
almost unchanged for any choice of the length from $40$ to $300$. 
In the same way, we can also estimate the values of $x_{\rm cross}$ for the 
other $V/t$ values. In Fig.~\ref{fig:fit_17_V195}\ (b), the estimated values of 
$x_{\mathrm{cross}}$ are plotted as a function of $V/t$. 
We find that the logarithmic correction is hardly present at $V \lesssim 1.8$ 
and the length scale $x_{\mathrm{cross}}$ increases rapidly in the vicinity of 
the CO insulating phase. Note that $x_{\rm cross} \to \infty$ at $V/t \to 2$.

\subsection{For finite $U$}

\begin{figure}[t]
\includegraphics[width=6.0cm]{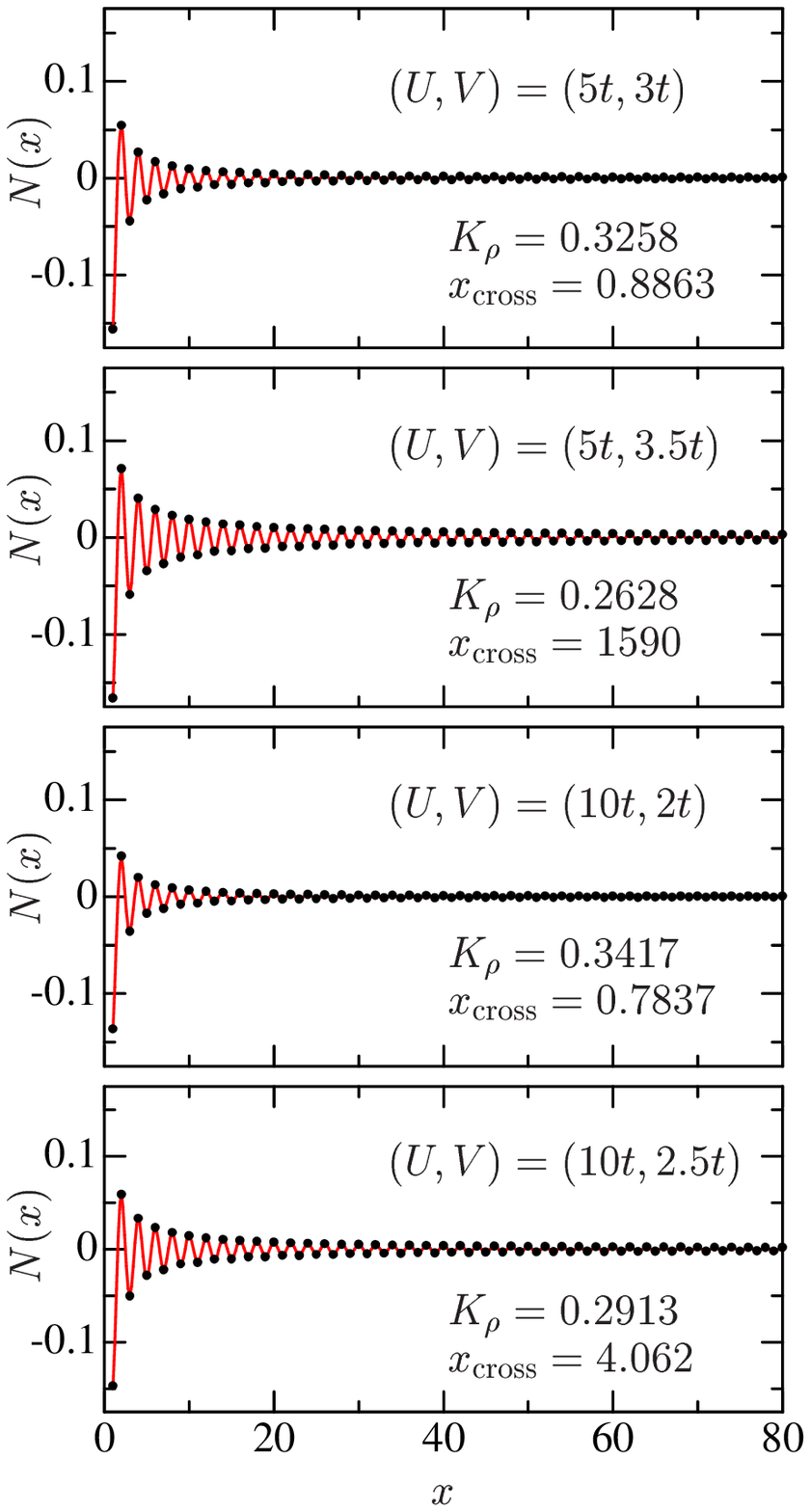}
\caption{
(Color online)
DMRG results of the correlation function $N(x)$ for several sets of $(U,V)$.  
The solid lines denote fitting curves with Eq.\ (\ref{eq:generalform}). The 
estimated values of $x_{\rm cross}$ and $K_\rho$ are also included.
\label{fig:log_correction}
 }
\end{figure}

We next turn to the case of $U \neq \infty$. In this case, the numerical results of 
the correlation function $N(x)$ can be fitted with the analytical form of 
Eq.\ (\ref{eq:generalform}). Differently from the case of $U = \infty$, there are 
five fitting parameters; namely, $\alpha$, $\alpha_\sigma$, $A_1$, $A_2$, and $K_\rho$. Of them, 
$K_\rho$ may be obtained very accurately with the DMRG method via the derivative 
of charge structure factor at $q=0$,
\begin{equation}
K_\rho=\frac{1}{2} \lim_{q \to 0} \langle n(q) n(-q) \rangle,
\end{equation}
with $q=2\pi/L$ and $n(q)=\sum_{l,s} e^{-iql}c_{l,s}^\dagger c_{l,s}$. Thus, we can reduce 
the fitting parameters from five to four. In Fig.~\ref{fig:log_correction}, we show 
the fitting results of the correlation function $N(x)$ near the boundary of the CO phase 
for $U=5t$ and $10t$ [the critical boundary has been estimated as 
$V_c \approx 3.70t$ ($2.76t$) for $U=5t$ ($10t$) in Ref.\ \onlinecite{Ejima2005}]. 
We can see that the DMRG data is in good agreement with the fitted line for all the 
parameter sets. From the obtained results of $x_{\mathrm{cross}}$, we find that 
the logarithmic correction appears for $(U,V)=(5t,3.5t)$ and $(10t,2.5t)$. Especially 
at $(U,V)=(5t,3.5t)$, the length scale is extremely large $x_{\mathrm{cross}} \approx 1600$; 
it allows us to crossly notice that this point is very close to the boundary of the CO phase. 
Meanwhile, the logarithmic correction is hardly present for $(U,V)=(5t,3t)$ and 
$(U,V)=(10t,2t)$. As a result, we confirm that the logarithmic correction is present 
also for $U \neq \infty$ and its length scale grows rapidly near the CO phase boundary. 

\begin{figure}[t]
\includegraphics[width=5.5cm]{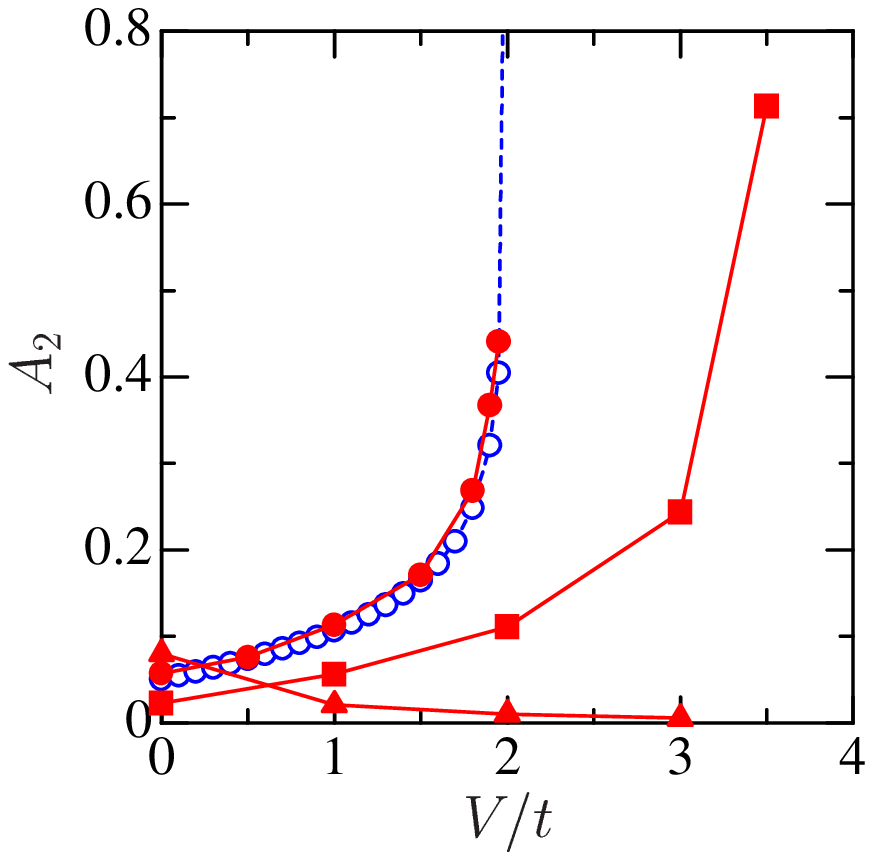}
\caption{
(Color online)
Correlation amplitude $A_2$ for $U=1$ (triangles), $5$ (squares), and $\infty$ (circles). 
Solid (dashed) lines denote the DMRG (exact) results.
\label{A2}
 }
\end{figure}

Finally, we discuss the correlation amplitudes, $A_1$ and $A_2$, of 
Eqs.\ (\ref{eq:generalformUinf}) and (\ref{eq:generalform}). Figure~\ref{A2} 
shows the DMRG results of the amplitude $A_2$ as a function of $V/t$ for 
several values of $U/t$. In the $U = \infty$ limit, we can see an excellent 
agreement between the DMRG and exact results. We also find a very sharp 
increase near $V=V_{\rm c}=2t$. For $U=5t$, the behavior of $A_2$ seems 
to be quite similar to that for $U = \infty$; while, the amplitude $A_1$ 
is rapidly decreased near $V=V_{\rm c}=3.70t$, e.g., $A_1 \lesssim 10^{-2}$ 
at $V \gtrsim 3$. Thus, the $2k_{\rm F}$ oscillating term would be negligible 
in the vicinity of the CO phase. When $U=t$, the amplitude $A_2$ decreases 
with increasing $V/t$ in reflecting that the $4k_{\rm F}$ fluctuation is 
not enhanced by $V$. We note that the two amplitudes $A_1$ and $A_2$ are 
rather small with the same order of magnitude for small $U$ and larger 
$V$ values.

\section{Summary}

We study the density-density correlation function in the TLL state for the 
1D extended Hubbard model at quarter filling. Based on the bosonization and 
RG techniques, we obtain the generalized analytical form of the correlation 
function which exhibits anomalous power-law behavior with logarithmic corrections 
near the phase boundary to the CO insulating state. Using the DMRG method, 
we confirm the appearance of the logarithmic corrections not only in the 
$U = \infty$ limit but also for finite $U$. Moreover, we find that the length 
scale of the corrections grows rapidly near the CO phase boundary.

\acknowledgments
 
The authors thank
   A.\ Furusaki and E.\ Orignac
for valuable discussions.

\appendix
\section{Derivation of the analytical form of the correlation function}

In this appendix, we derive the generalized form of the 
  correlation function 
  [Eqs.\ (\ref{eq:generalformUinf}) and (\ref{eq:generalform})]
 based on the RG approach.
In our derivation, 
we follow the formalism of the RG method developed in Ref.\ 
  \onlinecite{Giamarchi1989}.

The RG equations for the TLL parameter $K_\rho(l)$ and 
  the 1/4-filled umklapp scattering $G_\mathrm{u}(l)$ are given by
\cite{Giamarchi_book,Tsuchiizu2001}
\begin{subequations}
\begin{eqnarray}
\frac{d}{dl} K_\rho(l) &=& -2 G_\mathrm{u}^2(l) \, K_\rho^2(l),
\\
\frac{d}{dl} G_\mathrm{u}(l) &=& [2-8K_\rho(l)] G_\mathrm{u}(l),
\end{eqnarray}%
\label{eq:RGcoupling}%
\end{subequations}
where the initial values are estimated based on the perturbative
treatment in Ref.\ \onlinecite{Tsuchiizu2001}.
The TLL parameter in the low-energy effective theory 
  can be evaluated from the fixed point value of $K_\rho(l)$, i.e.,
  $K_\rho=K_\rho(\infty)$.
The correlation functions for the $2k_F$ and $4k_F$ oscillation parts,
  defined as
  $C_{2k_F}(x-x') \equiv
   2\langle \sin (2k_Fx+\phi_\rho(x))\sin (2k_Fx'+\phi_\rho(x'))\rangle/
    \cos 2k_F(x-x')$ and
  $C_{4k_F}(x-x') \equiv
   2\langle \cos (4k_Fx+\phi_\rho(x))\sin (2k_Fx'+\phi_\rho(x'))\rangle/
    \cos 2k_F(x-x')$, respectively, 
are given in the RG scheme by \cite{Giamarchi_book,Giamarchi1989}
\begin{subequations}
\begin{eqnarray}
C_{2k_F}(x) &=& \exp\left[-\int_0^{\ln (x/\alpha_0)} dl K_\rho(l)\right] ,
\\
C_{4k_F}(x) &=& \exp\left[-\int_0^{\ln (x/\alpha_0)} dl 
 \bigl(4K_\rho(l) - 2G_\mathrm{u}(l)\bigr)\right] ,
\nonumber \\
\end{eqnarray}%
\label{eq:RGcorrelation}%
\end{subequations}
where $\alpha_0$ is the short-distance cutoff.
The couplings $K_\rho(l)$ and $G_\mathrm{u}(l)$
  are determined by solving Eq.\ (\ref{eq:RGcoupling}).

From Eq.\ (\ref{eq:RGcoupling}), we find that 
  the fixed point values are given by
  $(K_\rho(\infty),G_\mathrm{u}(\infty))=(\frac{1}{4},0)$
   on the 
  phase boundary between the TLL  and CO states.
Near this phase boundary,  the TLL parameter can be expanded as 
 $K_\rho (l)=\frac{1}{4}+\frac{1}{4}G_\rho(l)$ 
  and we can treat $G_\rho(l)$  perturbatively.
Up to the second order in $G_\rho(l)$ and $G_\mathrm{u}(l)$, 
  the RG equations  (\ref{eq:RGcoupling}) are rewritten as
\begin{eqnarray}
\frac{d}{dl} G_\rho(l) = -2 G_\mathrm{u}^2(l)
, \quad
\frac{d}{dl} G_\mathrm{u}(l) = -2 G_\rho(l) G_\mathrm{u}(l).
\label{eq:RGcoupling2}%
\end{eqnarray}
In the case of $G_\rho(0) > |G_\mathrm{u}(0)|$,
  the umklapp scattering
   $G_\mathrm{u}(l)$ flows to zero, i.e., is irrelevant, and 
  $G_\rho(l)$ has a finite fixed point $G_\rho(\infty)\ge 0$. 
Thus the TLL parameter in the low-energy limit is given by
  $K_\rho=\frac{1}{4}+\frac{1}{4}G_\rho(\infty)$.
The explicit solutions of Eq.\ (\ref{eq:RGcoupling2}) are given by
\begin{subequations}
\begin{eqnarray}
G_\rho(l) 
&=& 
\frac{\theta}{2} \mathrm{coth} 
\left[
\theta l + \tanh^{-1} (\theta/2G_\rho(0))
\right],
\\
G_\mathrm{u}(l) 
&=& 
\frac{\theta}{2} \mathrm{cosech} 
\left[
\theta l + \tanh^{-1} (\theta/2G_\rho(0))
\right],
\end{eqnarray}%
\label{eq:RGsolution}%
\end{subequations}
where $\theta\equiv 2(G_\rho^2 - G_\mathrm{u}^2)^{1/2}$  is 
  a scaling invariant quantity.
Near the phase boundary, i.e., for small $\theta$,
 the umklapp scattering $G_\mathrm{u}(l)$ approaches to zero 
 very slowly as increasing $l$.
By substituting Eq.\ (\ref{eq:RGsolution}) into Eq.\
  (\ref{eq:RGcorrelation}),
  we obtain the analytical form of the correlation functions:
\begin{eqnarray}
C_{2k_F}(x)
&=& 
\left(\frac{\alpha_0}{x}\right)^{1/4+\theta/8}
\left(
\frac{1-d^{2\theta}}{1-(d\alpha_0/x)^{2\theta}}
\right)^{1/8},
\\
C_{4k_F}(x)
&=& 
\left(\frac{\alpha_0}{x}\right)^{1+\theta/2}
\left(
\frac{1+d^\theta}{1+(d\alpha_0/x)^\theta}
\right)^{3/2}
\nonumber \\ && {} \times
\left(
\frac{1-(d\alpha_0/x)^\theta}{1-d^\theta}
\right)^{1/2},
\end{eqnarray}
where $d$ is the nonuniversal quantity depending on the initial values
  of RG equations, defined by 
 $d\equiv \exp[-\theta^{-1}\tanh^{-1}(\theta/2G_\rho(0))]$.
In terms of $K_\rho$, 
 the parameter $\theta$ is given by  $\theta=(8K_\rho -2) $.
By defining $\alpha \equiv d\alpha_0$ 
we can derive Eqs.\ (\ref{eq:generalformUinf}) and
  (\ref{eq:generalform}).

\end{document}